\documentstyle{article}

\def\abstract#1{\vskip 7mm 
        \begin{center}{\large Abstract}\par \smallskip
                \begin{minipage}[c]{12cm}
                        \small #1
                \end{minipage}
        \end{center}
}
\def\title#1{\begin{center}{\Large\bf #1}\end{center}}
\def\author#1{\vskip 5mm \begin{center}{#1}\end{center}}
\def\address#1{\begin{center}{\it #1}\end{center}}
\makeatletter
\@ifundefined{lesssim}{}{}
\@ifundefined{gtrsim}{}{}
\def\vereq#1#2{\lower3pt\vbox{\baselineskip1.5pt \lineskip1.5pt
\ialign{$\m@th#1\hfill##\hfil$\crcr#2\crcr\sim\crcr}}}
\makeatother

\begin{document}

\title{%
  Accelerating Flat Universe and Analytic Study of the m-z Relation  \smallskip \\
  }
\author{%
  Takao Fukui\footnote{E-mail: tfukui@dokkyo.ac.jp}}
  \address{%
  Department of Language and Culture, Dokkyo University, \\
  Gakuencho, Soka, Saitama, 340-0042, Japan
}
\abstract{ An accelerating flat universe with a variable cosmological term is obtained in the Robertson-Walker metric. The variable cosmological term is defined by the correction terms of the metric tensor field. Simple solutions of the scale factor and the cosmological term are shown. In this model of the universe, the magnitude-redshift relation is analytically studied to see if the model reproduces the tendency of the present observational data. The equation of state parameter $\gamma$ is touched.}
\section{Introduction}
One of the probable explanations of the magnitude-redshift relation of type Ia supernovae might be that by an accelerated expansion of the universe. There seems to be two major candidates causing the acceleration. The one candidate is the vacuum energy of the cosmological constant which is left behind from the inflation era to the later eras. It now exceeds the matter density and is essential for the current universe to accelerate [1]. 

The other candidate is the vacuum energy varing with a scalar field $\phi$, i.e. the variable cosmological term. It is known as a scalar quintessence which exerts a negative pressure and is in this regard same as the cosmological constant [2]. In [3], the type Ia supernovae observations are analyzed in favour of the quintessence scenario. This scenario can be written in a scalar-tensor theory. Exact solutions in a generalized scalar-tensor theory, for the scale factor and the scalar field sa well are obtained by Fukui et al. [4] and is examined numeically in [5]. The variable cosmological term is shown there as a function of $\phi$ and $\phi_{,i}\phi^{,i}$.
 
The major difference between the two candidates is in the values of the equation of state parameter $\gamma$. In the present work,
the equation of state is  $p=(\gamma-1)\epsilon$. The cosmological constant strictly gives the value of $\gamma=0$. On the other hand, the value of the parameter in the quintessence is variable with $\phi$ and is not always exactly 0. It means that the parameters for the ordinary matter and for the quintessence are not identical. The reason why only the latter parameter is variable has not been explained yet. Is it to be constant as the former parameter is constant for each era? Or Is the former parameter to be variable too? 

 Recently Fukui [6] presented a new time varying cosmological term only in the metric tensor field. That is, the new term is defined by the correction terms of the metric tensor field. In the cosmological model, if the value of the parameter $\gamma$ is exactly $0$, the cosmological term is constant and the universe expands exponentially. If the value is less than $2/3$,  the matter behaves as the quintessence and the cosmological term varies as $t^{-2}$ . Because of the negative pressure due to the entity, the universe accelerates its expanding velocity in a power-law fashion. We present the solutions below in more general forms than those in [6].
 
Next we apply the new cosmological term to the accelerated expansion of the universe and study analytically its effect on the magnitude-redshift relation [1,3].

\section{Robertson-Walker Flat Universe}
The Lagrangian density is assumed here as follows by taking the metric tensor corrections into consideration;
\begin{equation}
L=\frac{c^4}{16\pi G}\left[R-2\lambda_nR^n+3n(n-1)\lambda_nR_{lm}R^{lm}\right]+L_m,
\end{equation}
where $\lambda_n$ is constant and $n$ takes the values of 0, 1 and 2. The 3rd term in parentheses appears as a quadratic correction in the case of $n=2$. Another quadratic term $R_{lmst}R^{lmst}$ can be ignored on the basis of the generalized Gauss-Bonnet invariant. The particular combination in  the coefficients of the second and third terms are determined by the G-B invarint. Here we make the assumption that the second and third terms together are the cosmological term.
\begin{equation}
\Lambda\equiv -\lambda_n\left\{R^n-\frac{3n(n-1)}{2}R_{lm}R^{lm}\right\}.
\end{equation}
The justification of this assumption is mentioned in [6].
Hereafter we study only the case of $n=1$ according to the analysis made also in [6].
The field equation derived from Eq.(1) can be solved for the scale factor $a(ct)$ and the cosmological term $\Lambda(ct)$, in the flat Robertson-Walker metric. The conventinonal relation, $\epsilon a^{3\gamma}=const.(\equiv \epsilon_{\gamma})$ can be employed because the conservation law of the energy momentum tensor is assured to be valid by the field equation. We quote the solutions below for convenience.\\
(a) $\gamma=0:$ The scale factor is obtained as follows;
\begin{equation}
a=a_0exp(\sqrt{P_0}t),
\end{equation}
where $P_0\equiv 8\pi G\epsilon_0/3(1-2\lambda_1)c^2$. The deceleration parameter is $q=-1$.\\
The cosmological term is obtained as follows;
\begin{equation}
\Lambda=\frac{32\pi G\epsilon_0\lambda_1}{c^4(1-2\lambda_1)}.
\end{equation}
During this era, the cosmological term is constant as usual.\\
 The energy densities and pressures are obtained as follows; 
\begin{eqnarray}
\epsilon&=&-p=\epsilon_0,\nonumber\\
\epsilon_{vac}&=&-p_{vac}=\frac{2\lambda_1\epsilon_0}{1-2\lambda_1}.
\end{eqnarray}
Both of them too are constant. From Eq.(5), $0<\lambda_1<1/2$.\\
(b) $\gamma\neq 0:$ The scale factor is shown here in a more general form than that in [6] as follows;
\begin{equation}
a=\left\{\frac{6\pi G\epsilon_\gamma \gamma^2}{(1-2\lambda_1)c^2}\right\}^{\frac{1}{3\gamma}}\left(t+\frac{a_*}{Q_\gamma}\right)^{\frac{2}{3\gamma}},
\end{equation}
where $Q_\gamma^2$ is defined as $8\pi G\epsilon_\gamma/3(1-2\lambda_1)c^2$ and thus is different for each $\gamma$. $a_*$ is  constant which is taken to be 0 in [6]. From Eqs.(3) and (6), an exponential expansion is restricted strictly to $\gamma=0$. Otherwise the universe expands in a power-law fashion. The deceleration parameter in (b) is $q=(3\gamma-2)/2$. Therefore the expansion of the universe accelerates when $\gamma<2/3$. From the observation $-1\leq\gamma-1\leq-0.93$ [7], which means that $\gamma$ is not strictly 0, the present scenario leads the expansion of the universe to accelerate not in an exponential manner but in a power-law fashion with $\gamma<2/3$.

The cosmological term is obtained as follows;
\begin{equation}
\Lambda=-\frac{4(3\gamma-4)\lambda_1}{3\gamma^2}\frac{1}{\{c(t+a_*/Q_\gamma)\}^2}.
\end{equation}
During these eras, the cosmological term decays as $t^{-2}$, though $\Lambda=0$ during the radiation era $\gamma=4/3$.
The energy densities and pressures are obtained as follows;
\begin{eqnarray}
p&=&(\gamma-1)\epsilon=(\gamma-1)\frac{(1-2\lambda_1)c^2}{6\pi G\gamma^2(t+a_*/Q_\gamma)^2}\nonumber\\
p_{vac}&=&(\gamma-1)\epsilon_{vac}=(\gamma-1)\frac{\lambda_1c^2}{3\pi G\gamma^2(t+a_*/Q_\gamma)^2}.
\end{eqnarray}
As is seen in Eqs.(5) and (8), both matter and vacuum share the same equation of state parameter $\gamma$.
 \section{Whole History of the Universe}
Here we assume that the universe evolves in terms of $\gamma$ [8], to the current accelerating universe as follows.

\begin{tabular}{c||c|c|c|c|c|c} \hline
era&big bang&inflation&radiation&matter&Inflection&present(quintessence)\\\hline
$\gamma$&$\gamma_b$&$0$&$4/3$&$1$&$2/3$&$\gamma_p<2/3$\\ \hline
$q$&$q_b$&$-1$&$1$&$1/2$&$0$&$q_p<0$\\ \hline
\end{tabular}\\

At each interface between eras, the junction conditions require that the scale factor $a(ct)$ and its derivative $\dot{a}(ct)$ be continuous. The conditions are satisfied at the following epochs.\\
(I) The transition epoch $t=t_{ir}$ between inflation ($\gamma=0$) and radiation ($\gamma=4/3$) is obtained as follows;
\begin{equation}
t_{ir}=\frac{1}{\sqrt{P_0}}\left(\frac{1}{2}\mp\sqrt{\frac{\epsilon_0}{\epsilon_{4/3}}}a_*\right).
\end{equation}
(II) The other transition epochs $(t_{rm}, t_{mI}, t_{Ip}) $ between $\gamma_1$ $(=4/3, 1, 2/3)$ and $\gamma_2$ $(=1, 2/3, \gamma_p<2/3)$ are obtained as follows;
\begin{equation}
t_{12}=\pm\frac{a_*}{\gamma_1-\gamma_2}\frac{1}{\sqrt{P_0}}\left(\gamma_2\sqrt{\frac{\epsilon_0}{\epsilon_{\gamma2}}}-\gamma_1\sqrt{\frac{\epsilon_0}{\epsilon_{\gamma1}}}~\right).
\end{equation}
The double sign in Eqs.(9) and (10) is owing to the sign of $Q_{\gamma}$ in Eq.(6). Here we include the inflection era as a transition from a deceleration to an acceleration. We might, however, ignore the era as an inflection point for a negligibly short period.
 
On the conditions, the solution for the scale factor can be employed in the application of the present model to observational cosmology.
\section{Magnitude-Redshift Relation}
The magnitude-redshift relation in a flat universe is given as follows;
\begin{equation}
m=M+5log{(1+z)a_R(r_R-r_E)}+K(z)+const.,
\end{equation}
here $K(z)$ is the K-correction which is a correction for the effects of the redshift of the spectrum on the apparent magnitude [9]. The correction is a purely instrumental effect [10], and is not studied in the present work because the aim of this work is to see only an effect due to an accelerating
expansion of the universe on the m-z relation. $(1+z)a_R(r_R-r_E)$ is the luminosity distance to a source of redshift $z$. The subscriptions $R$ and $E$ denote receiving in the  present era ($\gamma=\gamma_p)$ and emitting in the matter era ($\gamma=1$) respectively. The radial comoving distance $r_R-r_E$ can be obtained exactly by using the solution of the scale factor, Eq.(6) as follows;
\begin{eqnarray}
r_R-r_E&=&\int_{t_E}^{t_R}\frac{c}{a}dt=\int_{t_E}^{t_{mI}}\frac{c}{a_1}dt+
\int_{t_{mI}}^{t_{Ip}}\frac{c}{a_{2/3}}dt+\int_{t_{Ip}}^{t_R}\frac{c}{a_{\gamma_p}}dt\nonumber\\
&=&-3c\sqrt{\frac{(1-2\lambda_1)a_Rc^2}{6\pi G\epsilon_1}}\frac{1}{\sqrt{1+z}}
+3c\sqrt{\frac{(1-2\lambda_1)a_{mI}c^2}{6\pi G\epsilon_1}}\nonumber\\
&+&c\sqrt{\frac{3(1-2\lambda_1)c^2}{8\pi G\epsilon_{2/3}}}ln\frac{a_{Ip}}{a_{mI}}\nonumber\\
&+&\frac{3c}{3\gamma_p-2}\sqrt{\frac{(1-2\lambda_1)a_R^{3\gamma_p-2}
c^2}{6\pi G\epsilon_{\gamma_p}}}\left(1-\sqrt{\left(\frac{a_{Ip}}{a_R}\right)^{3\gamma_p-2}}~\right).
\end{eqnarray}
The third term might be small in its effect, or be ignored as is stated above. The last term is essential in the model of an accelerating universe.
For the case of the negligible inflection era, i.e. $a_{mI}\sim a_{Ip}(\equiv a_{mp})$, Eq.(12) becomes as follows by using the relation $\epsilon_1=\epsilon_{\gamma_p}a_{mp}^{3(1-\gamma_p)}$;
\begin{eqnarray}
r_R-r_E=\frac{2c}{a_R H_R}\left\{\frac{1}{3\gamma_p-2}-\sqrt{\left(\frac{a_{mp}}{a_R}\right)^{3(\gamma_p-1)}}\frac{1}{\sqrt{1+z}}+\frac{3(\gamma_p-1)}{3\gamma_p-2}\sqrt{\left(\frac{a_{mp}}{a_R}\right)^{3\gamma_p-2}}\right\},
\end{eqnarray}
where $H_R$, the Hubble parameter at the receiving time $t_R$, is given by the energy density $\epsilon_R=\epsilon_{\gamma_p}a_R^{-3\gamma_p}$ or $t_R$ as follows,
\begin{equation}
H_R=\frac{2}{3}\sqrt{\frac{6\pi G\epsilon_R}{(1-2\lambda_1)c^2}}=\frac{2}{3\gamma_p(t_R+a_*/Q_{\gamma_p})}.
\end{equation}
For the case of $\gamma_p=1$ (i.e. $t_{mI}=t_{Ip}=t_R$), Eq.(13) reduces to the following conventional equation;
\begin{equation}
r_R-r_E=\frac{2c}{a_RH_R}\left(1-\frac{1}{\sqrt{1+z}}\right).
\end{equation}
Since we study the objects with the redshift up to $z=1$, i.e. $a_R/a_E=1+z\leq2$, we can assume $a_R/a_{mp}\equiv(1+\delta)$ with an inequality $\delta<z\leq 1$, because $a_E<a_{mp}$. Then Eq.(13) can be approximated to  $O(\delta)$ as follows,
\begin{equation}
r_R-r_E\sim\frac{2c}{a_RH_R}\left(1-\frac{1}{\sqrt{1+z}}\right)\left\{1+\frac{3(1-\gamma_p)}{2}\delta\right\}.
\end{equation}
Again when $\gamma_p=1$, Eq.(16) reduces to Eq.(15). In the case of an accelerating universe where $\gamma_p<2/3$, the acceleration works for a larger magnitude which supports the observational data. The same tendency of the Hubble diagrams in [1,3] has been shown by Fukui in the case of the cosmological models with the cosmological constant containing the interacting matter and radiation [11]. The shaded plausible region in a (lambda-density) plane is shown there and is consistent with the tendensy [3] inferred from the recent observations.
\section{Comments}
By using Eqs.(9) and (10), the transition epoch $t_{Ip}$ is obtained as follows;
\begin{equation}
t_{Ip}=\frac{1}{(2-3\gamma_p)\sqrt{P_0}}\left\{\pm3\gamma_p\sqrt{\frac{\epsilon_0}{\epsilon_{\gamma_p}}}a_*-\sqrt{P_0}(-4t_{ir}+t_{rm}+t_{mI})-2\right\}.
\end{equation}
It is sure that $-4t_{ir}+t_{rm}+t_{mI}$ in Eq.(17) is positive.
Then, on a necessary condition for $t_{Ip}$ to be positive, the upper sign in Eq.(17) should be taken in the case of $a_*>0$, and the lower sign in the case of $a_*<0$. In either case, $a_*/Q_{\gamma}$ in Eq.(6) is not negative. 

Here, from $a\cong 0$ at $t=0$, we infer $a_*/Q_{\gamma_b}\cong 0$ in Eq.(6) only for the period $0(s)\leq t\leq 10^{-37}(s)$, the ``big bang" era. The inference might be acceptable when we notice that $a_*/Q_{\gamma}$ is different for each era in the present scenario [12]. Then the scale factor is connected smoothly between the big bang and the inflation at the following epoch;
\begin{equation}
t_{bi}=\frac{2}{3\gamma_b}\frac{1}{\sqrt{P_0}}.
\end{equation}
Since the ratio of the scale factors at $t_{ir}$ and $t_{bi}$  is obtained by Eqs.(3) and (18) as follows;
$$ \frac{a_{ir}}{a_{bi}}=exp\left\{\frac{2}{3\gamma_b}\left(\frac{t_{ir}}{t_{bi}}-1\right)\right\},$$
 we obtain $\gamma_b\sim 2/3$ for $t_{ir}/t_{bi}\sim100$ and $a_{ir}/a_{bi}\sim10^{43}$ [12]. Since $q_b\sim0$, the universe might have started its expansion with a constant speed.
 
In the present work, we applied the case of $n=1$ in Eq.(1) to one of the observables in cosmology, m-z relation and could show the observational tendency. Therefore the variable cosmological term originated in the correction terms of the metric tensor field [6] might be a possible alternative to that originated in the scalar-tensor field.

\end{document}